\documentclass[paper]{aa}
\usepackage{graphicx}
\usepackage{txfonts}
\def\1e{1E~1207.9+3945}
\def\ein{{\it Einstein}}
\def\sax{{\it Beppo}SAX}
\def\xmm{XMM-{\it Newton}}
\def\cha{{\it Chandra}}
\def\sw{{\it Swift}}
\def\eg{{\em e.g.\ }}
\def\nh{$N_{\rm{H}}$}

\def\den{$\times\,10^{\,20}$cm$^{-2}$}

\begin{document}
\title{The 26 year-long X-ray light curve and the X-ray spectrum of the BL~Lac Object 
       \1e in its brightest state}
\author{A.~Maselli\inst{1}, P.~Giommi\inst{2,3}, M.~Perri\inst{3}, R.~Nesci\inst{1}, 
       A.~Tramacere\inst{1}, F.~Massaro\inst{4}, M.~Capalbi\inst{3}}

\institute{Dipartimento di Fisica, Universit\`a La Sapienza, Piazzale A. Moro 2, 
           I-00185 Roma, Italy 
      \and Agenzia Spaziale Italiana, Unit\`a Osservazione dell'Universo, Viale Liegi 26, 
           I-00198 Roma, Italy
      \and ASI Science Data Center, ESRIN, Via G. Galilei, 
           I-00044 Frascati, Italy 
      \and Dipartimento di Fisica, Universit\`a Tor Vergata, Via della Ricerca Scientifica 1, 
           I-00133 Roma, Italy}

\offprints{alessandro.maselli@uniroma1.it} 

\date{Received 3 August 2007; Accepted 24 September 2007}

\titlerunning{\sw\,\,and multi-frequency observations of the BL~Lac object \1e}

\authorrunning{A.~Maselli et al.}

\abstract
{}
{We studied the temporal and spectral evolution of the synchrotron
emission from the high energy peaked BL~Lac object \1e.}
{Two recent observations have been performed by the \xmm\,\,and
\sw\,satellites; we carried out X-ray spectral analysis for both of
them, and photometry in optical-ultraviolet filters for the
\sw\,\,one. Combining the results thus obtained with archival data we
built the long-term X-ray light curve, spanning a time interval of
26~years, and the Spectral Energy Distribution (SED) of this source.}
{The light curve shows a large flux increasing, about a factor of six,
in a time interval of a few years. After reaching its maximum in
coincidence with the \xmm\,\,pointing in December~2000 the flux
decreased in later years, as revealed by \sw\,. The very good
statistics available in the 0.5-10 keV \xmm\,\,X-ray spectrum points
out a highly significant deviation from a single power law. A
log-parabolic model with a best fit curvature parameter of 0.25 and a
peak energy at $\sim$~1~keV describes well the spectral shape of the
synchrotron emission. The simultaneous fit of \sw\,\,UVOT and XRT data
provides a milder curvature ($b\sim0.1$) and a peak at higher energies
($\sim15$~keV), suggesting a different state of source activity. In
both cases UVOT data support the scenario of a single synchrotron
emission component extending from the optical/UV to the X-ray band.}
{New X-ray observations are important to monitor the temporal and
spectral evolution of the source; new generation $\gamma$-ray
telescopes like AGILE and GLAST could for the first time detect its
inverse Compton emission.}


\maketitle


\section{Introduction}

BL~Lac objects are thought to be radio-loud Active Galactic Nuclei
(AGNs) observed in a direction very close to the axis of a
relativistic jet outflowing from the inner nuclear region
(\cite{Urry95}). This interpretation could explain most of the
characteristics of these sources like compact and flat-spectrum radio
emission, superluminal motion revealed by VLBI imaging, high and
variable radio and optical polarization, non-thermal continuum
emission extending from radio to $\gamma$-ray frequencies, an almost
featureless optical spectrum and the fast variability at all
frequencies.

BL~Lac objects are generally characterized by a double bump structure
in the broad band Spectral Energy Distribution (SED). The low
frequency bump is attributed to synchrotron radiation emitted by
relativistic electrons in the jet; inverse Compton scattering by the
same electron population on the synchrotron radiation is thought to be
at the origin of the high frequency bump. The peak of the first bump
may vary in a rather wide range of frequencies: from the IR/optical
band for the low energy peaked BL~Lacs (LBLs) to the UV/X-ray band for
the high energy peaked BL~Lacs (HBLs) (\cite{Giommi94},
\cite{Padovani}).

\1e, also named BZB J1210+3929 in the recent Multifrequency Catalogue
of Blazars (\cite{BZcat}) is one of the X-ray selected BL~Lacertae
objects of the \ein\,\,Medium Sensitivity Survey (\cite{EMSS}). It was
discovered as a serendipitous source located about five arc-minutes
north of one of the most intensively studied AGNs, the bright Seyfert
galaxy NGC~4151. For this reason \1e has been observed on many
occasions by all the imaging X-ray instruments that have operated
since the \ein\,\,observatory. Despite its relatively high redshift
(z=0.615) HST was able to detect the bright (M$_{\rm R}$\,=\,$-$24.4)
host galaxy which is of elliptical type (\cite{Scarpa00}).

In this paper we report the long-term X-ray light curve which spans
over 26 years. We also report and compare the spectral analysis of the
most recent observations carried out by the \xmm\,\,(\cite{jansen})
and \sw\,\,(\cite{gehrels}) satellites; we discuss the possibility to
model the synchrotron emission of this source with a single
log-parabolic model. The peak of the synchrotron component lies in the
X-ray band, and for this reason the source can safely be classified as
an HBL. We finally report the Spectral Energy Distribution (SED) of
\1e compiled from non-simultaneous multi-frequency archival data.

\section{The long-term X-ray light curve}

Because of its spatial proximity to the well known Seyfert Galaxy
NGC~4151, \1e has been observed by a large number of X-ray instruments
on board several astronomical spacecrafts. We collected all available
historical X-ray data from literature and we also added more recent
results from \xmm, \cha\, and \sw\,\,observations.

Most of the available data were accessed through the ASI Science Data
Center (ASDC) on-line services (\texttt{www.asdc.asi.it}). Data
relative to \ein\,\,IPC were taken from the 2E~catalog (\cite{Harris})
while those relative to \ein\,\,HRI come from the Einstein Observatory
HRI source list (\cite{giacconi}). Data from ROSAT~PSPC were derived
from WGACAT2 catalog (\cite{WGA}); data from ROSAT~HRI were obtained
from the BMW catalog (\cite{BMW}). We completed these informations
with those relative to EXOSAT~CMA reported in Giommi
et~al.~(1990). All these measurements, expressed in cts/s, have been
opportunely converted to monochromatic fluxes at 1~keV; each value has
also been corrected for Galactic absorption adopting \nh~=~2.0~\den
\,, a value estimated from 21~cm measurements along the line of sight
(\cite{Dickey90}).

\begin{figure}[thbf]
\includegraphics[width=7.5cm, angle=-90]{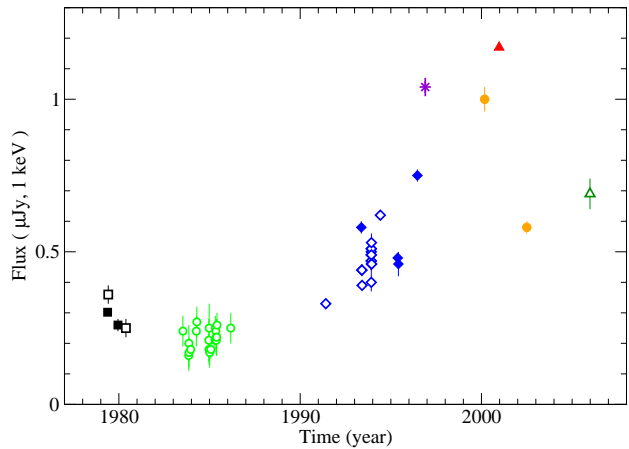}
\caption{The long-term X-ray light curve of \1e. Starting from the
earliest observations we plotted data from \ein\,\,IPC (filled
squares) and HRI (empty squares), EXOSAT~CMA (empty circles),
ROSAT~PSPC (empty diamonds) and HRI (filled diamonds), \sax\,\,MECS
(star), \cha\,\,(filled circles), \xmm\,\,EPIC MOS1 (filled triangle)
and \sw\,\,XRT (empty triangle) instruments.}
\label{lc}
\end{figure}

The MECS instrument on board the \sax\,\,satellite observed \1e in
July and December~1996. As Cusumano et al.~(2001) found no relevant
change in the source flux over this time interval, they performed a
spectral analysis summing the four available observations in a single
spectral file; we report the flux at 1~keV derived from the results of
their spectral analysis. As regards the most recent observations (from
\xmm, \cha\,\,and the latest by \sw) we performed data reduction and
derived the flux at 1~keV directly from the spectral fit. The light
curve thus obtained spans a very long time interval, more than 25
years, and is plotted in Fig.~\ref{lc}.

%
%
%
%
%
\begin{table*}[bhpt]
\caption{Basic informations on \xmm\,\,and \sw\,\,observations. To
distinguish the two different \xmm\,\,exposures, both performed on 22
December 2000, we indicate with I the longer exposure, and with II the
other one.}
\label{log}
\begin{center}
\begin{tabular}{ccccccc}
\hline
\hline
          &            &             &                 &             &          &            \\
Satellite & Instrument &  Sequence   &     Readout     &    Date     & Exposure & Net counts \\
          &            &   number    &      mode       &             &   (s)    &            \\
          &            &             &                 &             &          &            \\
\hline
          &            &             &                 &             &          &            \\
\xmm      & EPIC-MOS1  & 0112830201  &  Full Window    & 22 Dec 2000 &   59361  &   34236    \\
          &            &    (I)      & Medium Filter   &             &          &            \\
          &            &             &                 &             &          &            \\
\xmm      & EPIC-MOS1  & 0112830501  &  Full Window    & 22 Dec 2000 &   22092  &   12418    \\
          &            &    (II)     & Medium Filter   &             &          &            \\
          &            &             &                 &             &          &            \\
\sw       &    XRT     & 00035463001 & Photon Counting & 26 Dec 2005 &   10368  &    1218    \\
          &            &             &                 &             &          &            \\
\hline

\multicolumn{7}{c} { }
\end{tabular}
\end{center}
\end{table*}
%

The light curve is characterized by the following behaviour: the flux
remained low (0.2-0.3 $\mu$Jy) for a $\simeq$ 10 year-long time
interval during the '80s, and then it increased during '90s to a high
state, maintaining a level around 1 $\mu$Jy for about four years. The
highest flux was detected by \xmm\,\,by the end of 2000, and it is
about a factor of six higher than the mini\-mum. After reaching its
peak, the source luminosity decreased in the latest years, with the
flux at the epoch of the most recent \sw\,\,pointing (December 2005)
at a level comparable to the one detected by ROSAT ten years earlier,
about three times higher than the minimum. Superposed on this trend,
flux variations of smaller amplitude ($\sim$ 0.1 $\mu$Jy) and much
shorter time scales (even 1-2 days) have been put in evidence thanks to
the very close sequence of observations (one day frequency) in
December 1993 by the ROSAT satellite.

\section{XMM-Newton observations}
\label{tre}

\1e was observed by \xmm\,\,on 22~December~2000, by means of the
EPIC-PN (\cite{xmm3}) and EPIC-MOS (\cite{xmm4}) CCD~came\-ras,
operating in ``full window" frame with medium filters. We
restricted our analysis to the EPIC-MOS data only because for other
HBL~sources the estimates of the spectral parameters were found
consistent with those obtained from \sax~and \sw~observations
(Tramacere et al.~2007, Massaro et al.~2007). We checked both MOS1 and
MOS2 data, and after verifying the consistency between results we
reported only those relative to MOS1. The pointing was splitted in
two intervals, one significantly longer than the other; details are
reported in Table~\ref{log}. Both these observations were reduced
adopting standard criteria and following the ``\textit{User's Guide to
the \xmm\,\,Science Analysis System (SAS)}'' (Issue~3.1) (\cite{xmm1})
and ``\textit{The \xmm\,\,ABC~Guide}" (vers.~2.01)
(\cite{xmm2}). Particular care was dedicated to the solar flares
subtraction, as described in Tramacere et al.~(2007a).

Photons for spectral analysis were extracted from a circular region
with a radius of $40\arcsec$. The background spectrum was extracted
from a circular region of size comparable to the source region, in a
place where visible sources were not present. The Photon
Redistribution Matrix and the Ancillary Region File were created for
each observation, by using \textsc{rmfgen} and \textsc{arfgen} tasks
of SAS.

To ensure the validity of Gaussian statistics we grouped data so that
each new bin included at least 40~counts. Spectral analysis was
carried out using XSPEC~11.3.2 (\cite{arnaud}) in the 0.5--10 keV
restricted energy range to avoid possible residual calibration
uncertainties. We analyzed separately the two observations and we
found best fit parameters consistent within the errors, indicating
that the source spectrum did not change significantly on such a short
time scale; we thus applied a simultaneous fit of the two spectra.

First, we fitted the two data sets using a single power law model:
\begin{equation}
F(E) = K ~ E^{-a} ~~~~{\rm ph\;cm}^{-2}\;{\rm s}^{-1}\;{\rm keV}^{-1}
\end{equation}
with the hydrogen equivalent column density fixed at the Galactic
value: we found $\chi^2_r/d.o.f.=1.23/376$ and residuals showing
evidence of an intrinsic curvature.
%
%
\begin{figure}[hftb]
\includegraphics[width=5.8cm, angle=-90]{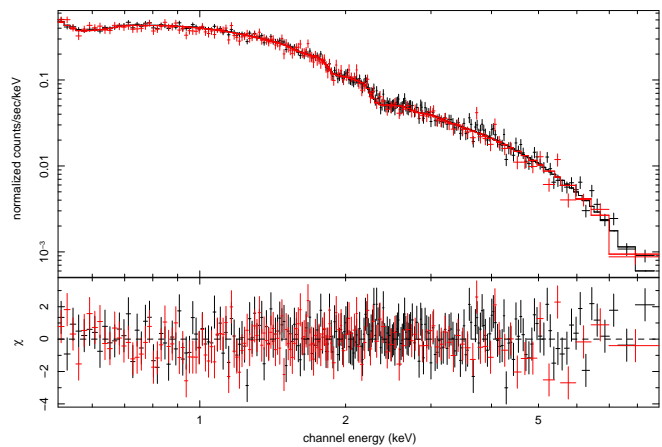}
\caption{The simultaneous fit of the \xmm\,\,observations I and II
adopting a log-parabolic law in the 0.5--10 keV energy range; the
\nh\, parameter is fixed to the known Galactic value.}
\label{sim_pwlw2_col_cl.ps}
\end{figure}
%
%
Repeating the fit with the same model and leaving \nh\,free to vary,
we found a fit statistical improvement ($\chi^2_r/d.o.f.=1.07/375$)
but the absorbing column density value resulted almost a factor of
three (\nh = 5.3 \den) higher than the Galactic one. All parameter
values relative to the spectral analysis are reported in
Table~\ref{tab1}. We also used a broken power-law model with
Galactic \nh~and found a good fit ($\chi^2_r/d.o.f.=0.98/374$) with
spectral photon indices a$_1=2.01 \pm 0.01$ and a$_2=2.34\pm0.04$
before and after the energy break E$_b=2.2^{+0.1}_{-0.3}$ keV,
respectively.
%
%
\begin{table*}[thbp]
\caption{Best fit spectral parameters for \xmm\,\,and \sw\,\,observations
derived from XSPEC analysis. Statistical errors are taken at 1$\sigma$
confidence level. For each satellite we reported the results obtained
adopting the power law model with fixed and free \nh\/, and the
log-parabolic model. We also added results derived fixing the
curvature parameter to the value ($b=0.18$) obtained from fitting non
simultaneous \sw\,\,UVOT and \xmm\,\,fluxes (see Sect.\ref{Discussion}).}
\label{tab1}
\begin{center}
\begin{tabular}{cccccccc}
\hline 
\hline 
 & & & & & & & \\
Satellite & \nh\/ & $K$ & $a$ & $b$ & $E_p$ & $F_{2-10}$ & $\chi^2_r$/d.o.f.\\
 & (10$^{20}$ cm$^{-2}$) & (10$^{-3}$ ph cm$^{-2}$ s$^{-1}$ keV$^{-1}$) & & & (keV) & (10$^{-12}$ erg cm$^{-2}$ s$^{-1}$) & \\
 & & & & & & \\    
\hline 
 & & & & & & \\
XMM & 2.0 (fixed) & 1.74$\pm$0.01 & 2.10$\pm$0.01 & & & 3.8 & 1.23/376 \\
XMM & 5.3$\pm$0.4 & 1.92$\pm$0.03 & 2.21$\pm$0.02 & & & 3.6 & 1.07/375 \\
XMM & 2.0 (fixed) & 1.76$\pm$0.01 & 2.00$\pm$0.01 & 0.25$\pm$0.03 & 1.0 & 3.5 & 1.01/375 \\
\hline 
XMM & 2.0 (fixed) & 1.76$\pm$0.01 & 2.03$\pm$0.01 & 0.18 (fixed) & 0.8 & 3.5 & 1.02/376 \\
\hline 
\sw & 2.0 (fixed) & 0.96$\pm$0.03 & 1.91$\pm$0.04 & & & 2.8 & 1.34/52 \\
\sw & 7.9$\pm$1.7 & 1.22$\pm$0.08 & 2.16$\pm$0.09 & & & 2.5 & 1.09/51 \\
\sw & 2.0 (fixed) & 1.04$\pm$0.04 & 1.77$\pm$0.06 & 0.43$\pm$0.13 & 1.9 & 2.4 & 1.12/51 \\
 & & & & & & & \\
\hline

\multicolumn{8}{c} { }
\end{tabular}
\end{center}
\end{table*}
%
%

An alternative model that provides a good fit is given by
the log-parabolic model (\cite{massaro04}):
\begin{equation}
F(E) = K ~ (E/E_0)^{-(a + b Log (E/E_0))} ~~~~{\rm ph\;cm}^{-2}\;{\rm s}^{-1}\;{\rm keV}^{-1}
\end{equation}
where $a$ is the spectral index (given by the log-derivative) at
$E_0$~=1~keV and $b$ measures the spectral curvature. This spectral
model is able to represent an intrinsic curved spectra with the
addition of just one parameter to the single power law; it has been
successfully applied to modelling the SED of other HBL sources like
\eg Mrk~421 (\cite{massaro04}), Mrk~501 (\cite{massaro04b}),
PKS~0548-322 (\cite{matteo}) and also a sample of HBLs detected at Tev
energies (\cite{TramaTeV}). Moreover, it let estimate in a very simple
way other interesting quantities, like the energy of the SED peak
given by the following expression:
\begin{equation}
E_p = E_0~10^{~(2-a)/2b} ~~~~{\rm  keV}.
\end{equation}
The log-parabolic model provides for \1e a very good $\chi^2_r$ value
($\chi^2_r/d.o.f.$ = 1.01/375) and a curvature parameter
($b=0.25\pm0.03$) consistent with typical values obtained in the
spectral analysis of other HBL sources (all parameter values are
reported in Table~\ref{tab1}). Figure~\ref{sim_pwlw2_col_cl.ps}
reports the spectrum and the residuals of the analysis in the whole
range (0.5--10 keV).

To estimate the significance of the improvement of the power law model
with free \nh\, and the log-parabolic model with respect to the power
law model with Galactic \nh\/, we applied an F-test. In both cases we
obtained very low values ($5.0\times10^{-13}$ and $8.6\times10^{-18}$,
respectively) so that the two alternative models are significantly
better than the power law with a Galactic \nh\/ value. The
log-parabolic model has the additional advantage that it does not
require any intrinsic absorption, originating in the innermost nuclear
environment or in the host galaxy, to justify the spectral curvature.

\section{Swift observation}
\label{swift}

\1e was observed in the field of NGC~4151 on 26~December~2005 with
both the UVOT (UltraViolet-Optical Telescope, \cite{roming}) and XRT
(X-Ray Telescope, \cite{burrows}) instruments.

The XRT observation was carried out using the most sensitive Photon
Counting readout mode (see \cite{hill} for a description of readout
modes). Data were reduced with the XRTDAS software package
(version~2.1.2) developed at the ASDC (Capalbi et al.~2005) and
distributed within the HEASOFT~6.3 package by the NASA High Energy
Astrophysics Archive Research Center (HEASARC).

We used the \textsc{ximage} task to detect the source count rate and
the centroid, and to choose a nearby source-free region in order to
extract the background spectrum. The source count rate was $\sim 0.2$
cts/s, a value not high enough ($\geq 0.5$ cts/s) to determine photon
pile-up occurrence. Spectral data were then extracted in a circular
region within 20 pixels radius (1 pixel = $2.36\arcsec$) from the
centroid; the background spectrum was estimated in a circular region
of 50 pixels radius.

We then used the \textsc{xrtproducts} task to obtain all the files
necessary to the spectral analysis; the exposure map was taken into
account to correct for vignetting, CCD hot and damaged pixels. In the
spectral file, instrumental channels were combined to include in each
new energy bin at least 20 counts. The spectral analysis was carried
out in the 0.3--10~keV energy range using XSPEC~11.3.2.

Following the same criteria described in Sect.~\ref{tre} we first fit
the data using a single power law model, both fixing the hydrogen
equivalent column density to the Galactic value and leaving it free to
vary (see Table~\ref{tab1}). In the second case we obtained, as
expected, a better $\chi^2_r$ value ($\chi^2_r/d.o.f.$ = 1.09/51 vs
1.34/52) but an \nh\/ value almost a factor of four (\nh = 7.9 \den)
higher than the Galactic one. Again we fit the log-parabolic model
with Galactic \nh\/ to data, obtaining a comparable improvement
($\chi^2_r/d.o.f.$ = 1.12/51) with no extra absorption needed (see
Table~\ref{tab1}). The F-test gives a probability of
$2.62\times10^{-3}$ ($\sim$ 3 $\sigma$) that this improvement is due
to chance.

\begin{figure}[bhft]
\includegraphics[width=5.8cm, angle=-90]{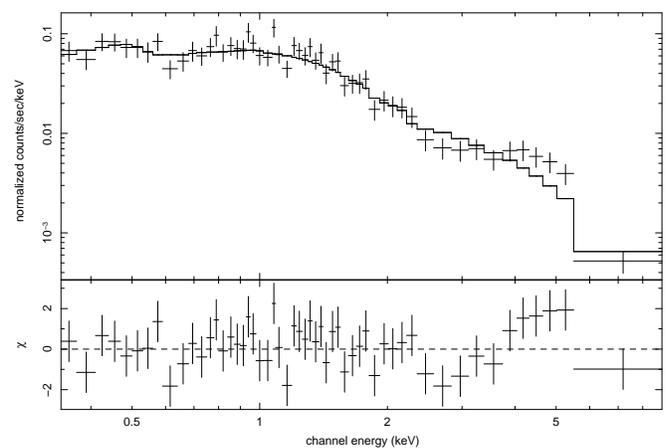}
\caption{The spectral analysis of the \sw\,\,XRT observation in the
0.3--10 keV energy range adopting a log-parabolic model and fixing the
\nh\, parameter to the known Galactic value.}
\label{biascorr_009_gr20_parab.ps}
\end{figure}

The spectrum and the residuals are plotted in
Fig.~\ref{biascorr_009_gr20_parab.ps}. The distinctive feature of this
plot is a dip around 3~keV followed by an apparent raising at higher
energies. New \sw\,\,observations with an exposure time longer than 10~ks
would help reducing the fluctuations characteristic of this spectrum.
%
%
%
%
\subsection{UVOT data reduction}
\label{quattrouno}

The UVOT instrument obtained series of images in each of the
lenticular filters $V, B, U, W1, M2, W2$. The data analy\-sis was
performed using dedicated tasks included in the HEASOFT~6.3
package. Each series was summed up with the \textsc{uvotimsum} task to
obtain a single frame in each filter; photometry was then performed
using the \textsc{uvotsource} task.
%
%
\begin{figure}[t]
\vspace{1.0cm}
\includegraphics[width=9cm]{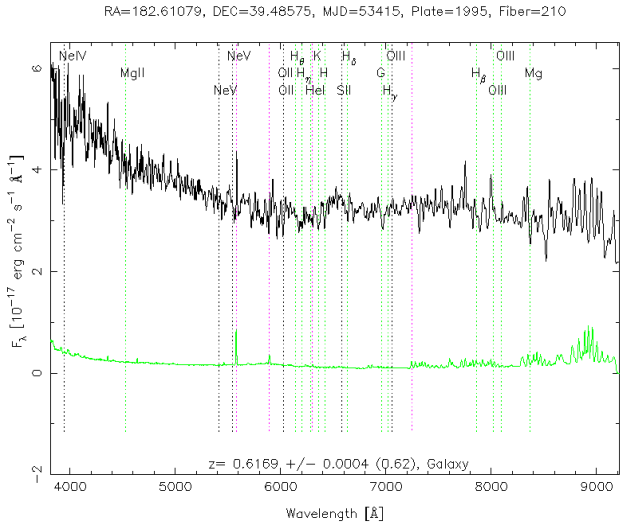}
\caption{The optical spectrum from SDSS, with the evident change in
the spectral slope at $\lambda \simeq 6000$ \AA.}
\label{spettrosloan}
\end{figure}
%

Following the UVOT Team prescriptions (\cite{uvotman}) we chose an
aperture radius of 2$\arcsec$ for doing photometry, independently of
the image filter; then we made an aperture correction to transfer the
obtained magnitude to the standard photometry aperture that was used
to obtain the photometric zero point, which is 6$\arcsec$ for $V, B,
U$ filters and 12$\arcsec$ for the $W1, M2, W2$ filters. To achieve
this purpose we selected a sample of stellar sources with a high
signal-to-noise ratio and we calculated the difference between the
magnitudes measured with the two different radii. After deriving a
mean value for the aperture correction in each filter we finally
obtained the corrected magnitudes for \1e.

The background region was chosen in the form of an annulus around the
extraction region, with inner and outer radii taken opportunely in
order to avoid the inclusion of spurious sources, particularly
numerous in $UV$ filter frames; typical values are respectively
18$\arcsec$ and 23$\arcsec$. Then the obtained magnitudes were
de-reddened using the value of $E(B-V) = 0.03$ mag (\cite{Schlegel98})
and adopting the extinction curve given in \cite{Seaton79} with
R$_V = 3.2$, and finally converted to specific fluxes; results
are reported in Table~\ref{tab_uvot}.
%
%
\begin{table}[bhtp]
\caption{\1e de-reddened magnitudes in the UVOT six filters, and
corresponding specific fluxes.}
\label{tab_uvot}
\begin{center}
\begin{tabular}{ccc}
\hline
\hline
       &                  &                   \\
Filter &    Magnitude     &   Specific flux   \\
       &      (mag)       &     ($\mu$Jy)     \\
       &                  &                   \\
\hline
       &                  &                   \\
  V    & 19.76 $\pm$ 0.19 &   39.2 $\pm$ 6.9  \\
  B    & 20.30 $\pm$ 0.24 &   28.9 $\pm$ 6.4  \\
  U    & 18.90 $\pm$ 0.09 &   36.4 $\pm$ 3.0  \\
  W1   & 18.85 $\pm$ 0.08 &   27.3 $\pm$ 2.0  \\
  M2   & 18.87 $\pm$ 0.09 &   26.0 $\pm$ 2.1  \\
  W2   & 18.98 $\pm$ 0.08 &   24.6 $\pm$ 1.8  \\
       &                  &                   \\
\hline

\multicolumn{3}{c} { }
\end{tabular}
\end{center}
\end{table}
%
%

\section{Optical spectrum and host galaxy contribution}

\1e is present in the POSS-I and POSS-II plates at a very faint level,
very close to the plate sensiti\-vity limits (B and R $\simeq 20$),
and no appreciable variation is apparent between the two epochs. The
source is also present in the Sloan Digital Sky Survey (SDSS) and has
been observed on 17 February 2004; also in this case no significant
variation can be found comparing magnitudes in g and r filters to
magnitudes in the POSS B and R filters. The SDSS optical spectrum
(\texttt{http://cas.sdss.org}) is reported in Fig.~\ref{spettrosloan}:
it does not show evidence of prominent lines and presents a slope
change around $ \lambda \simeq 6000$~\AA. This variation can be
interpreted as the sign of the combined BL~Lac and host galaxy
emission: while the first one dominates at higher frequencies, its
contribute is overwhelmed by the host galaxy at higher wavelengths.
%
%
\begin{figure}[ht]
\includegraphics[width=7cm, angle=-90]{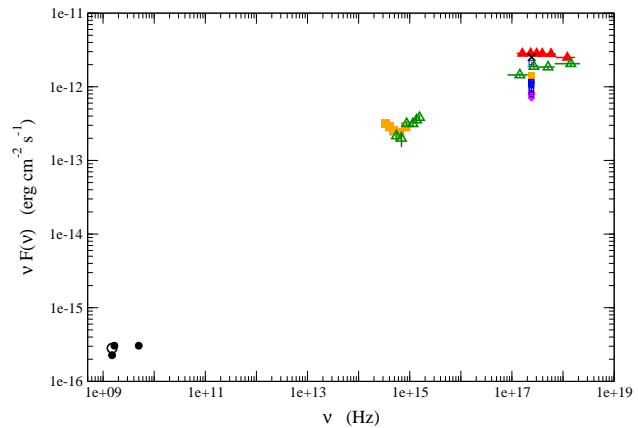}
\caption{The Spectral Energy Distribution of \1e. Filled black circles
represent simultaneous radio data taken at the VLA (\cite{Stocke})
while the empty circle corresponds to the FIRST pointing; the nearly
coincident value in NVSS has not been plotted for simplicity. Filled
orange squares representing SDSS data overlap \sw\,\,UVOT empty
triangles (dark green) in the optical region of the spectrum. At 1~keV
energy, \ein\,\,IPC are plotted in magenta (filled diamonds) and ROSAT
PSPC in blue (empty squares); the orange filled square corresponds to
the detection in the ROSAT All Sky Bright Source Catalog (1RXS)
(\cite{1RXS}) while the black cross corresponds to \sax. The same
symbol as UVOT has been used to plot \sw\,\,XRT data; red filled
triangles refer to the \xmm\,\,observation~I.}
\label{SED}
\end{figure}
%
%
%
%
\begin{figure*}[ht]
\hspace{2.5cm}
\includegraphics[width=10cm, angle=-90]{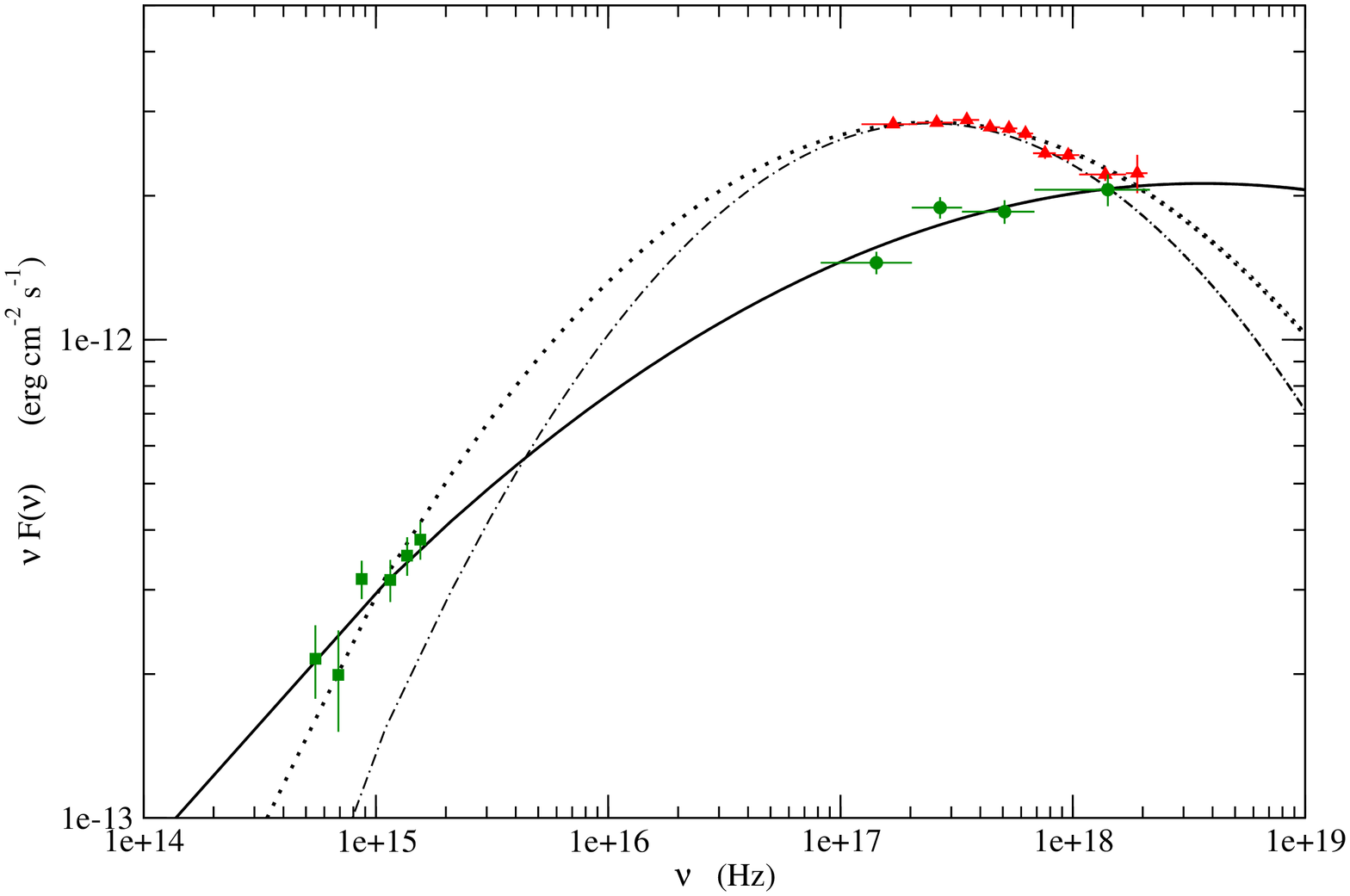}
\caption{An optical to X-ray SED of \1e including \sw\,\,UVOT
(squares), \sw\,\,XRT (circles) and \xmm\,\,MOS1 (triangles) data
sets. Note that the X-ray data have been further rebinned in
XSPEC with respect to the criteria followed in the spectral
analysis and described in Sect.~\ref{tre} and Sect.~\ref{swift}. A
log-parabolic model has been fitted to different data sets: solid line
($b\sim0.1$) represents the best fit to \sw\,\,UVOT and XRT
simultaneous fluxes; dot-dashed line ($b\sim0.25$) represents the best
fit to the \xmm\,\,data points, extended to a wider frequency
interval; dotted line ($b\sim0.18$) instead represents the best fit to
non simultaneous UVOT and \xmm\,\,(observation~I) data sets.}
\label{combo}
\end{figure*}
%
%
Host galaxy contribution clearly emerges also from the inspection of
the Spectral Energy Distribution (SED) in the frequency interval
$10^{14}<\nu<10^{15}$ Hz (see Fig.~\ref{SED}). To verify the agreement
between the informations in the optical spectrum and in the SED we
graphically estimated the slope variation in the two different
plots. At this purpose we divided the SDSS spectrum at 6000~\AA \, in
two intervals and graphically estimated the slope change on both
sides. Analogously, we took SDSS points in the SED and splitted them
in two sets, corresponding to the IR-optical (z, i, r) and optical-UV
(g, u) filters; we traced a line interpolating the two different sets
and estimated their slope and the relative variation. The same value
of $\Delta \alpha \simeq 1.2$ that we found in both cases supports the
interpretation that the host galaxy is mostly responsible for the
infrared emission visible in the SED.

\section{Discussion}
\label{Discussion}

Adding to multi-frequency archival data the results of our analysis we
built the radio to X-ray SED of \1e (Fig.\ref{SED}) which follows the
characteristic trend of extreme HBL sources (\cite{Padovani}) with the
synchrotron emission covering the entire frequency interval and
peaking well into the X-ray band. The inverse Compton component is not
detected and therefore is expected at higher energies.

The \sw\,\,satellite is sensitive both in the X-ray and in the
optical-ultraviolet band. This provides the opportunity to test if the
Spectral Energy Distribution of a source like \1e can be explained by
a single synchrotron component, or if multiple emission components are
present. For this purpose we first fitted a log-parabolic model to
simultaneous \sw\,\,XRT and UVOT fluxes (solid line in
Fig.~\ref{combo}) and obtained a rather low value for the curvature
parameter ($b\sim0.1$) in contrast with the one obtained by fitting
XRT data only, as reported in Table~\ref{tab1}; the energy peak lies
at about 15~keV or more. Comparing these results with those from \xmm,
\sw\,\,evidently caught \1e in a different state of activity,
characterized by a milder curvature and a peak of the synchrotron
component shifted to higher energies. Anyway firm conclusions cannot
be drawn due to a relatively poor statistics and to the fact that the
found $E_p$ value likely lies at energies higher than 10~keV, outside
the range of \sw\,\,XRT.

Aware of the limits of using non simultaneous observations for a
variable source, yet we veri\-fied if \sw\,\,UVOT and \xmm\,\,MOS1
data are compatible with a log-parabolic model. We took points
representing XSPEC model of observation~I and fit them on a wider
frequency interval (dot-dashed line in Fig.\,\ref{combo}): UVOT points
are sistematically above the extrapolation of the log-parabolic
model. We took care to compare this result with the one obtained by
fitting directly X-ray rebinned data and obtained an almost coincident
result. Then we fitted both data sets with a log-parabola (dotted line
in Fig.\,\ref{combo}) and estimated the curvature parameter: we
obtained $b=0.18\pm0.01$, a value at about 2$\sigma$ of the one
derived fitting only \xmm\,\,data (see Table \ref{tab1}). At this
point we tested the log-parabolic model with $b$ fixed at 0.18 and
found $\chi^2_r/d.o.f. = 1.02/376$, a value practically coincident
with the one obtained leaving $b$ free to vary, with only a small
difference in the $a$ parameter. This result encouraged us in
concluding that a single synchrotron component is in a good agreement
with the emission observed.

\1e is certainly worth monitoring in the next years. New observations
in the X-ray band would add essential information to the light curve
behaviour: if any kind of regularity in flux increasing and fading
should emerge, it would be possible to put physical constraints to
establish the nature of the mechanism responsible for these
variations. Moreover, observations with instruments which extend to
higher energies than XRT, like those on board \textit{Suzaku}, would
extend the known Spectral Energy Distribution and would allow us to
obtain a better parametrization of the spectral curvature. Finally,
\1e may be an interesting target for AGILE and the forthcoming GLAST
$\gamma$-ray mission which could detect for the first time the inverse
Compton component of this HBL source.

\begin{acknowledgements}

We wish to thank Enrico Massaro for useful discussions and helpful
comments, Gino Tosti and Simonetta Puccetti for hints in
\sw~UVOT and \textit{Chandra} data analysis respectively. The authors
acknowledge the financial support by the Italian Space Agency (ASI)
for ASDC and for the Physics Department of Universit\`a di Roma
``\textit{La Sapienza}''.

\end{acknowledgements}


\end{document}